\documentclass[conference]{IEEEtran}
\IEEEoverridecommandlockouts
\usepackage{cite}
\usepackage{amsmath,amssymb,amsfonts}
\usepackage{algorithmic}
\usepackage{graphicx}
\usepackage{textcomp}
\usepackage{xcolor}
\def\BibTeX{{\rm B\kern-.05em{\sc i\kern-.025em b}\kern-.08em
    T\kern-.1667em\lower.7ex\hbox{E}\kern-.125emX}}

\usepackage{hhline}
\usepackage{subcaption}
\usepackage{url}
\usepackage{color}
\usepackage{balance}
\usepackage{multirow}

\begin{document}


\title{Inplace knowledge distillation with teacher assistant for improved training of flexible deep neural networks\\
\thanks{The present work has been supported by European Union’s Horizon 2020 research and innovation program under grant number 951911 - AI4Media.}
}

\author{\IEEEauthorblockN{Alexey Ozerov}
\IEEEauthorblockA{\textit{InterDigital R\&D France} \\
Cesson-S\'evign\'e, France \\
alexey.ozerov@interdigital.com}
\and
\IEEEauthorblockN{Ngoc Q. K. Duong}
\IEEEauthorblockA{\textit{InterDigital R\&D France} \\
Cesson-S\'evign\'e, France \\
quang-khanh-ngoc.duong@interdigital.com}
}

\maketitle
\begin{abstract}
Deep neural networks (DNNs) have achieved great success in various machine learning tasks. However, most existing powerful DNN models are computationally expensive and memory demanding, hindering their deployment in devices with low memory and computational resources or in applications with strict latency requirements.
Thus, several resource-adaptable or flexible approaches were recently proposed that train at the same time a big model and several resource-specific sub-models.
Inplace knowledge distillation (IPKD) became a popular method to train those models and consists in distilling the knowledge from a larger model ({\it teacher}) to all other sub-models ({\it students}).
In this work a novel generic training method called {\it IPKD with teacher assistant (IPKD-TA)} is introduced, where sub-models themselves become {\it teacher assistants} teaching smaller sub-models. \\
We evaluated the proposed IPKD-TA training method using two state-of-the-art flexible models (MSDNet and Slimmable MobileNet-V1) with two popular image classification benchmarks (CIFAR-10 and CIFAR-100). Our results demonstrate that the IPKD-TA is on par with the existing state of the art while improving it in most cases.
\end{abstract}
\begin{IEEEkeywords}
Deep Neural Networks, Flexible Models, Inplace Knowledge Distillation with Teacher Assistant
\end{IEEEkeywords}
\section{Introduction}
\label{sec:intro}
Deep neural networks (DNNs) have achieved state-of-the-art results in many machine learning applications in the areas such as computer vision \cite{Huang2017, Hu2018}, speech recognition \cite{Amodei2016} and natural language processing \cite{Devlin2018}. Most of actual architectures are trained for specific tasks and have fixed complexity and performance at the inference. Although it is established that introducing more parameters often improves the accuracy of a model, bigger models are computationally and memory-wise too expensive to be deployed on the consumer electronics (CE) or internet of things (IoT) devices which have limited capacities (computational and memory resources).

One way to solve this problem consists in using model compression techniques targeting at the same time to reduce the model size and to accelerate it at the inference \cite{Cheng2018}. Most popular model compression methods are based on parameter quantization \cite{Vanhoucke2011} and pruning \cite{LeCun1990}, low-rank factorization \cite{Denton2014}, and knowledge distillation (KD) \cite{Hinton2015}. Quantization consists in simply quantizing the weights, while pruning consists in removing non-important weights \cite{LeCun1990} or entire convolutional channels \cite{Liu2017} (structural pruning) according to some criterion. Low-rank factorization-based techniques use matrix/tensor structural decompositions to approximate matrices/tensors of weights. In this work particular attention will be paid to KD-based methods that consist in compressing a big so-called {\it teacher} model by distilling its knowledge to a smaller (compressed) so-called {\it student} model. The KD itself is achieved during the training of student model by replacing or completing the ground truth labels with their teacher model probabilistic predictions. Model compression schemes, while very efficient, allow producing just one compressed model at a time that has fixed memory, run-time and performance characteristics. However, in many cases the available resources of the device on which the model will be executed may not be known in advance, and yet on the same device those resources might vary in time due to other processes. As such, models allowing for an on-demand trade-off between resources and performance, without any re-training or fine-tuning, become of great interest. 

Several resource-adaptable models/frameworks, here referred to as {\it flexible models}, that allow for such on-demand resources/performance trade-off have recently emerged \cite{Huang2018, Yu2019, Yu2019a, Cai2019, Ruiz2019, Ruiz2020, Guerra2020}. At a very high level of abstraction, a flexible model is constituted of several models (one per resources/performance operating point) that are embedded one into another like Matryoshka dolls with strong parameters sharing: one largest model and several sub-models. At the inference a suitable sub-model corresponding to the available resources may be instantly extracted and deployed, and, thanks to the strong parameters sharing, the full model may be efficiently transmitted and stored, as compared to a dummy solution of training several independent models (one per resources/performance operating point). Without loss of generality we here consider flexible DNNs applied to classification tasks, and, in particular, to image classification. Among most recent and efficient flexible models, we may mention the following ones. Multi-scale dense network (MSDNet) \cite{Huang2018} is a particular architecture with early-exits (classifications), where the flexibility is achieved by stopping computation at any desired classifier. Slimmable \cite{Yu2019} and universally slimmable \cite{Yu2019a} networks represent a general framework allowing for an instantaneous slimming \cite{Liu2017} (or structural pruning) of a single network into different sub-networks that are all trained jointly. One for all framework \cite{Cai2019} is based on the same principles as slimmable networks, while introducing more variability in sub-model's design, including elastic resolution, kernel size, depth and width. Ruiz and Verbee introduced convolutional neural mixture models \cite{Ruiz2019} and hierarchical neural ensembles \cite{Ruiz2020}, where the flexibility is achieved by selecting respective sub-mixtures or sub-ensembles. Finally, switchable precision neural networks \cite{Guerra2020} allow for a flexibility thanks to a possibility of on-demand switching between different levels of network weight quantization.

The most straightforward way to learn flexible models is to train all sub-models jointly from the annotated data \cite{Huang2018, Yu2019}. A more efficient training scheme called {\it inplace knowledge distillation (IPKD)} was introduced in \cite{Yu2019a} (then re-used in \cite{Cai2019, Ruiz2019, Ruiz2020, Guerra2020}), where the training is again joint, though only the biggest model is fully trained from the data, while all other sub-models are distilled from the biggest one. The new term {\it inplace} refers to the fact that there is a strong parameters sharing between the sub-models. However, in an alternative work on KD \cite{Mirzadeh2019}, that is not related to flexible models, it was noted that when the gap (in size) between the teacher and student models is large, the KD might be less efficient. To overcome this drawback the authors of \cite{Mirzadeh2019} introduce an in-between model, a so-called {\it teacher assistant}, that first learns from the teacher and then teaches the student.

In this work we build on the idea of KD with teacher assistant \cite{Mirzadeh2019} to improve the IPKD training of flexible models. Indeed, in IPKD all sub-models are distilled from the biggest one, and thus for smaller sub-models the teacher-student gaps are large. To overcome this issue, we introduce {\it IPKD with one teacher assistant (IPKD-TA-1)} and {\it IPKD with multiple teacher assistants (IPKD-TA-M)}, where each sub-model is distilled from the larger sub-model next to it (one teacher assistant) or from all the larger sub-models (multiple teacher assistants), respectively. Our proposal is general and applicable for training any flexible model, where IPKD is applicable. We investigate the effectiveness of the proposed approach in case of MSDNet \cite{Huang2018} and Slimmable \cite{Yu2019, Yu2019a} MobileNet-V1 \cite{Howard2017} models on two standard image classification benchmarks: CIFAR-10 and CIFAR-100 \cite{Krizhevsky2009}. Note that, to our best knowledge, even the IPKD approach was not yet explored for MSDNet model, possibly because MSDNet \cite{Huang2018} was published before IPKD was introduced in \cite{Yu2019a}. Our contributions may be summarized as: $(i)$ introducing novel IPKD-TA-1 and IPKD-TA-M approaches for improved training of general flexible models, $(ii)$ experimental investigation of IPKD approach for MSDNet, and $(iii)$ experimental investigation of proposed IPKD-TA-1 and IPKD-TA-M approaches for MSDNet and Slimmable MobileNet-V1.

The rest of this paper is organized as follows. Related work and necessary background are described in Section~\ref{sec:sota}. IPKD-TA-1 and IPKD-TA-M approaches for improved flexible models training are introduced in Section~\ref{sec:intro}. Section~\ref{sec:experiment} is devoted to experiments and conclusions are drawn in Section~\ref{sec:conclusion}.

\section{Related work and background}
\label{sec:sota}

\subsection{Knowledge distillation}

Knowledge distillation (KD) \cite{Hinton2015} is a general technique allowing to compress a big pre-trained model or model ensemble called {\it teacher} into a smaller {\it student} model. It is shown in \cite{Hinton2015} that using KD allows often for better performance than simply training the student model from the same data via supervised learning.

Let us consider a classification problem with $C$ classes. Let $C$-dimensional vectors $a_s$ and $a_t$ be the logits (the inputs to the final softmax) of the teacher and student networks, respectively. In classical supervised learning the student model is usually learned by optimizing the cross-entropy loss:~\footnote{Throughout the paper and without loss of generality, all the losses are expressed for just one data sample, and they should be averaged over the corresponding batch for a final implementation.}
\begin{equation}
    {\mathcal L}_{CE} (a_s, y_r) = {\mathcal H} \left( {\rm softmax} (a_s), y_r \right),
    \label{eq::cross_entr_loss}
\end{equation}
where $y_r$ is a $C$-dimensional hot vector encoding of the ground truth label, and ${\mathcal H} (x, z) = - \sum_{c=1}^C x_c \log (z_c)$ is the cross-entropy.

In KD framework \cite{Hinton2015} an additional term distilling the knowledge from the teacher is considered
\begin{equation}
    {\mathcal L}_{KD} (a_s, a_t) = \tau^2 {\mathcal D}_{KL} \left( {\rm softmax} (a_s / \tau), {\rm softmax} (a_t / \tau) \right),
    \label{eq::kd_loss}
\end{equation}
where ${\mathcal D}_{KL} (x, z) = \sum_{c=1}^C x_c \log \left( x_c / z_c \right)$ is the Kullback-Leibler (KL) divergence,~\footnote{According to \cite{Hinton2015} cross-entropy may be used as well in KD term \eqref{eq::kd_loss} instead of the KL divergence.} and hyperparameter $\tau$ referred to temperature is introduced to put additional control on softening of signal arising from the output of the teacher model. 

The final loss for training student model combines both the supervised learning loss and the KD loss as:
\begin{equation}
    {\mathcal L}_{KD}^{student} = (1 - \lambda) {\mathcal L}_{CE} (a_s, y_r) + \lambda {\mathcal L}_{KD} (a_s, a_t),
    \label{eq::kd_student_loss}
\end{equation}
where $\lambda \in [0, 1]$ is a constant hyperparameter weighing the contribution of each term.

\subsection{Knowledge distillation with teacher assistant}
\label{sec::kd_teacher_assistant}

It was remarked in \cite{Mirzadeh2019} that when the gap in model size between the teacher and the student is big enough, distilling the knowledge directly from the teacher might be sub-optimal. To overcome this issue the authors of \cite{Mirzadeh2019} introduced an intermediate {\it teacher assistant} model that is first learned by the teacher and then teaches the student. This approach has been shown \cite{Mirzadeh2019} to lead to a better student model performance.

The KD with teacher assistant concept may be easily understood intuitively. Indeed, an university professor may not be an ideal teacher for primary school children. However, the following usual way is efficient: a university professor teaches a school teacher who then teaches primary school children.

\subsection{Flexible DNNs}

We here describe briefly the two flexible models we consider in this work: MSDNet \cite{Huang2018} and Slimmable networks \cite{Yu2019, Yu2019a}. Though, there are many other flexible models \cite{Cai2019, Ruiz2019, Ruiz2020, Guerra2020} to which our proposed approach is applicable.

\subsubsection{Models}

{\bf MSDNet} \cite{Huang2018} is a densely-connected convolutional neural network (CNN) proposed for image classification. It has a two-dimensional (scale and depth) architecture and is divided along the depth dimension into several blocks. An early-exit classifier is implemented at the end of each block, and the flexibility is achieved by a possibility to stop the computation at any desired classifier.

{\bf Slimmable} networks \cite{Yu2019, Yu2019a} is a general framework applicable to CNNs that are suitable for slimming (a particular structural pruning) \cite{Liu2017}, and we here investigate it in case of MobileNet-V1 architecture \cite{Howard2017}. Sub-networks of a slimmable network are slimmed versions of the full network obtained by removing entire convolutional channels so as to have different widths, \emph{e.g.,} via scaling the full network width by 0.25, 0.5, 0.75, and 1.0. All sub-networks share their parameters, except for the batch normalization \cite{Ioffe2015} statistics.

\subsubsection{Conventional training}

Let flexible DNN include $n$ sub-models enumerated in the order of their sizes (the largest network is indexed by $n$).
Let vector $a_s[i]$ or $a_t[i]$ be the logits of the $i$-th model ($i = 1, \ldots, n$).~\footnote{Though the vectors $a_s[i]$ and $a_t[i]$ represent the same quantity, we distinguish between their notations to indicate within corresponding criteria whether a vector is in the role of a student ($a_s[i]$: the corresponding model parameters are optimized) or of a teacher ($a_t[i]$: the vector is just used as an input).}
The most conventional way to train flexible DNN consists in a supervised joint learning of all sub-networks as:
\begin{equation}
    {\mathcal L}^{flex} = \sum_{i=1}^n {\mathcal L}_{CE} (a_s[i], y_r),
    \label{eq::conv_flex_train_loss}
\end{equation}
with ${\mathcal L}_{CE} (\cdot, \cdot)$ specified in \eqref{eq::cross_entr_loss}.

\subsection{Inplace knowledge distillation for Flexible DNNs}

It was proposed in \cite{Yu2019a} to still train the sub-models jointly, but not all of them in a completely supervised manner: the largest $n$-th model is trained in a supervised way, while all other sub-models are distilled from the largest model. The resulting inplace knowledge distillation (IPKD) optimization loss writes
\begin{multline}
    {\mathcal L}^{flex}_{IPKD} = {\mathcal L}_{CE} (a_s[n], y_r) + (1 - \lambda) \sum_{i=1}^{n-1} {\mathcal L}_{CE} (a_s[i], y_r) + \\ + \lambda \sum_{i=1}^{n-1} {\mathcal L}_{KD} (a_s[i], a_t[n]),
    \label{eq::loss_ipkd}
\end{multline}
with ${\mathcal L}_{KD} (\cdot, \cdot)$ specified in \eqref{eq::kd_loss}.
A new adverb {\it inplace} in IPKD refers to the fact that the KD is now performed jointly with a strong parameters sharing between the sub-models.
It was shown in \cite{Yu2019a} for Slimmable networks that the IPKD training outperforms the conventional training \eqref{eq::conv_flex_train_loss}.
The IPKD strategy was then adopted in \cite{Cai2019, Ruiz2019, Ruiz2020, Guerra2020}.

\section{Proposed approaches}
\label{sec:intro}

We build our proposed training approaches on the idea of KD with teacher assistant (Sec.~\ref{sec::kd_teacher_assistant}), though introducing it within IPKD training of flexible DNNs. Indeed, all previous approaches \cite{Yu2019, Yu2019a, Cai2019, Ruiz2019, Ruiz2020, Guerra2020} are using IPKD without teacher assistant, \emph{i.e.,} by always distilling the knowledge from the largest model.

\subsection{IPKD with one teacher assistant}

We first introduce the IPKD with one teacher assistant (IPKD-TA-1), where each sub-model is taught by another sub-model that is just next to it from the top. This leads to the following re-formulation of IPKD loss \eqref{eq::loss_ipkd}:
\begin{multline}
    {\mathcal L}^{flex}_{IPKD-TA-1} = {\mathcal L}_{CE} (a_s[n], y_r) + (1 - \lambda) \sum_{i=1}^{n-1} {\mathcal L}_{CE} (a_s[i], y_r) + \\ + \lambda \sum_{i=1}^{n-1} {\mathcal L}_{KD} (a_s[i], a_t[i+1]).
    \label{eq::loss_ipkd-ta-1}
\end{multline}

\subsection{IPKD with multiple teacher assistants}

Another strategy we introduce and investigate is the IPKD with multiple teacher assistants (IPKD-TA-M), where each sub-model is taught by all the larger sub-models \footnote{In general each sub-model can be taught by a subset of larger sub-models.}. This is achieved by writing the corresponding loss as follows:
\begin{multline}
    {\mathcal L}^{flex}_{IPKD-TA-M} = {\mathcal L}_{CE} (a_s[n], y_r) + (1 - \lambda) \sum_{i=1}^{n-1} {\mathcal L}_{CE} (a_s[i], y_r) + \\ + \lambda \sum_{i=1}^{n-1} \frac{1}{n-i} \sum_{j=i+1}^{n} {\mathcal L}_{KD} (a_s[i], a_t[j]),
    \label{eq::loss_ipkd-ta-m}
\end{multline}
where the weights $\frac{1}{n-i}$ (such that $\sum_{j=i+1}^{n} \frac{1}{n-i} = 1$) are introduced in order to re-balance the impacts of the teacher assistants, since the number of teacher assistants varies from one sub-model to another. 

\section{Experiments}
\label{sec:experiment}

\begin{figure*}[ht]
\begin{subfigure}{.5\textwidth}
  \centering
  \includegraphics[width=.8\linewidth]{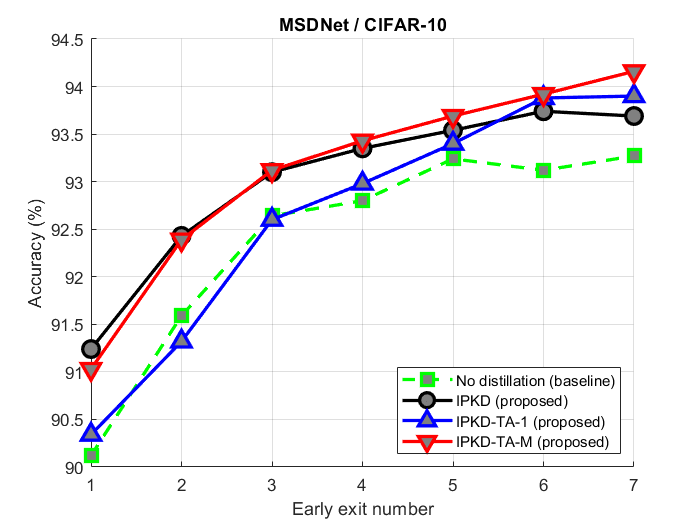}  
\end{subfigure}
\begin{subfigure}{.5\textwidth}
  \centering
  \includegraphics[width=.8\linewidth]{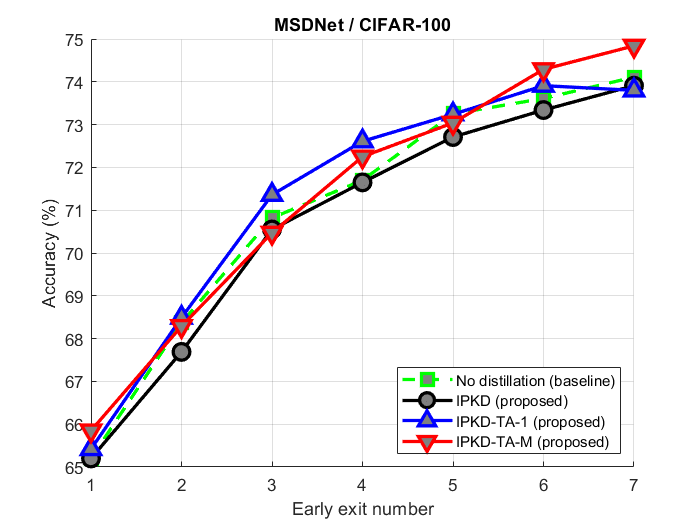}  
\end{subfigure}
\caption{Results in terms of classification accuracy for MSDNet on CIFAR-10 (left) and CIFAR-100 (right).}
\label{fig:MSDNet_results}
\end{figure*}

\begin{table*}[ht]
  \begin{center}
    \begin{scriptsize}
    \begin{tabular}{l||c|c|c|c||c|c|c|c||c|c|c|c}
      \hhline{=========}
      Dataset & \multicolumn{4}{c||}{CIFAR-10} & \multicolumn{4}{c}{CIFAR-100} \\
      \hhline{-||----||----}
      \multirow{2}{*}{Method} & No dist. & IPKD & IPKD-TA-1 & IPKD-TA-M & No dist. & IPKD & IPKD-TA-1 & IPKD-TA-M \\ 
             & (baseline) & (proposed) & (proposed) & (proposed) & (baseline) & (proposed) & (proposed) & (proposed) \\ 
      \hhline{=========}
        Exit 1 & 90.12 & \bf 91.24 & 90.34 & 91.03 & 65.20 & 65.20 & 65.42 & \bf 65.85 \\
        Exit 2 & 91.59 & \bf 92.43 & 91.32 & 92.39 & 68.37 & 67.69 & \bf 68.49 & 68.29 \\
        Exit 3 & 92.64 & 93.10 & 92.60 & \bf 93.12 & 70.82 & 70.55 & \bf 71.36 & 70.47 \\
        Exit 4 & 92.80 & 93.35 & 92.98 & \bf 93.43 & 71.70 & 71.65 & \bf 72.61 & 72.25 \\
        Exit 5 & 93.24 & 93.54 & 93.40 & \bf 93.69 & \bf 73.25 & 72.71 & 73.24 & 73.04 \\
        Exit 6 & 93.12 & 93.74 & 93.88 & \bf 93.92 & 73.61 & 73.34 & 73.91 & \bf 74.29 \\
        Exit 7 & 93.27 & 93.69 & 93.90 & \bf 94.16 & 74.11 & 73.91 & 73.80 & \bf 74.84 \\
      \hhline{-||----||----}
        Avg    & 92.39 & 93.01 & 92.63 & \bf 93.10 & 71.00 & 70.72 & 71.26 & \bf 71.29 \\
      \hhline{=========}
    \end{tabular}
    \end{scriptsize}
    \caption{Detailed results in terms of classification accuracy (\%) for MSDNet (best performance for each exit is in bold).}
    \label{tab:MSDNet_results}
  \end{center}
\end{table*}

\begin{figure*}[h!]
\begin{subfigure}{.5\textwidth}
  \centering
  \includegraphics[width=.8\linewidth]{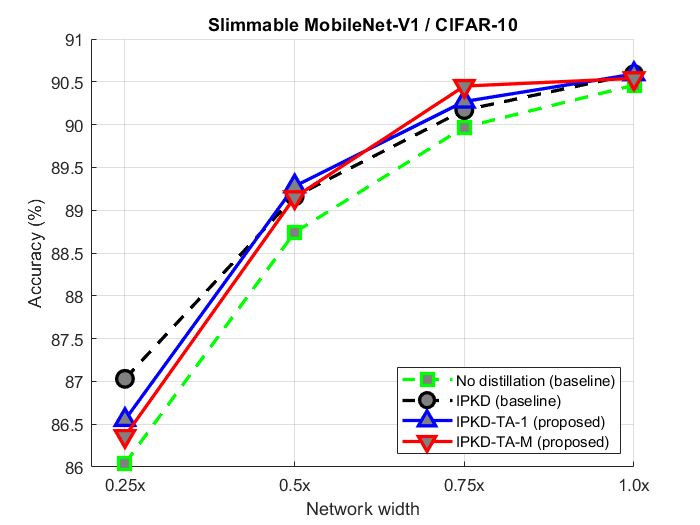}  
\end{subfigure}
\begin{subfigure}{.5\textwidth}
  \centering
  \includegraphics[width=.8\linewidth]{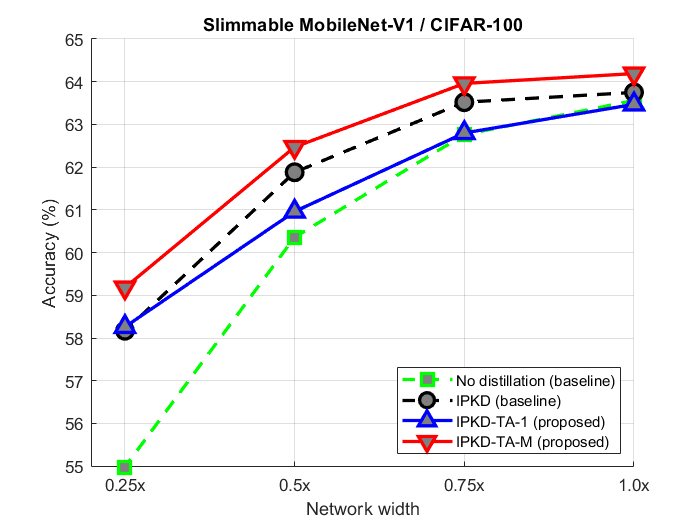}  
\end{subfigure}
\caption{Results in terms of classification accuracy for Slimmable MobileNet-V1 on CIFAR-10 (left) and CIFAR-100 (right).}
\label{fig:Slimmable_results}
\end{figure*}

\begin{table*}[h!]
  \begin{center}
    \begin{scriptsize}
    \begin{tabular}{l||c|c|c|c||c|c|c|c}
      \hhline{=========}
      Dataset & \multicolumn{4}{c||}{CIFAR-10} & \multicolumn{4}{c}{CIFAR-100}   \\
      \hhline{-||----||----}
      \multirow{2}{*}{Method} & No dist. & IPKD & IPKD-TA-1 & IPKD-TA-M & No dist. & IPKD & IPKD-TA-1 & IPKD-TA-M \\ 
             & (baseline) & (baseline) & (proposed) & (proposed) & (baseline) & (baseline) & (proposed) & (proposed) \\ 
      \hhline{=========}
        Switch 1 ($\times$ 0.25) & 90.46 & \bf 90.59 & \bf 90.59 & 90.54 & 63.56 & 63.75 & 63.47 & \bf 64.19 \\
        Switch 2 ($\times$ 0.5)  & 89.97 & 90.17 & 90.27 & \bf 90.45 & 62.76 & 63.52 & 62.80 & \bf 63.96 \\
        Switch 3 ($\times$ 0.75) & 88.74 & 89.16 & \bf 89.28 & 89.15 & 60.36 & 61.88 & 60.96 & \bf 62.47 \\
        Switch 4 ($\times$ 1.0)  & 86.04 & \bf 87.03 & 86.55 & 86.36 & 54.98 & 58.17 & 58.26 & \bf 59.18 \\
      \hhline{-||----||----}
        Avg    & 88.80 & \bf 89.24 & 89.17 & 89.13 & 60.42 & 61.83 & 61.37 & \bf 62.45 \\
      \hhline{=========}
    \end{tabular}
    \end{scriptsize}
    \caption{Detailed results in terms of classification accuracy (\%) for Slimmable MobileNet-V1 (best performance for each switch is in bold).}
    \label{tab:Slimmable_results}
  \end{center}
\end{table*}

\subsection{Datasets}

We evaluate the purposed approaches on two following popular image classification benchmarks. CIFAR-10 \cite{Krizhevsky2009} consists of 60k color images 32 $\times$ 32 split into 10 classes. CIFAR-100  \cite{Krizhevsky2009} consists of the same images as CIFAR-10, though they are split onto 100 classes. 

\subsection{Experimental setup}

We investigate the proposed approach on two different flexible models: MSDNet \cite{Huang2018} and Slimmable \cite{Yu2019, Yu2019a} MobileNet-V1 \cite{Howard2017}. Our approach and all the baselines are implemented based on the corresponding implementations and model architectures available at \url{https://github.com/kalviny/MSDNet-PyTorch} and \url{https://github.com/JiahuiYu/slimmable_networks}, respectively. Since the original Slimmable model implementation was only for ImageNet dataset (and not CIFAR-10 or CIFAR-100), some parameters, incuding stride, were adapted as in the following library: \url{https://github.com/weiaicunzai/pytorch-cifar100}.

All models are trained for 300 and 200 epochs for MSDNet and Slimmable MobileNet-V1, respectively, with an initial learning rate of 0.1 using the stochastic gradient descent (SGD) optimizer. We have chosen temperature  hyperparameter $\tau = 5$ and $\tau = 1$ in \eqref{eq::kd_loss} for MSDNet and Slimmable MobileNet-V1, respectively. Note also that for Slimmable MobileNet-V1 we used cross-entropy loss instead of the KL divergence in \eqref{eq::kd_loss}, since the same choice was done in the baseline implementation \cite{Yu2019, Yu2019a}. Penalty factor $\lambda$ in \eqref{eq::loss_ipkd}, \eqref{eq::loss_ipkd-ta-1} and \eqref{eq::loss_ipkd-ta-m} was set to $\lambda = 0.8$. 

The performance of each flexible DNN sub-model is measured in terms of classification accuracy \cite{Huang2018, Yu2019}. For \mbox{MSDNet} we measure the performance for each exit, and consider training without distillation as a baseline, since even IPKD was not yet applied for MSDNet. For Slimmable MobileNet-V1 we measure the performance for each of four switches ($\times$ [0.25, 0.5, 0,75, 1.0]), and consider the IPKD training with no distillation as a baseline. 
Following \cite{Yu2019, Yu2019a}, we report for each experiment the results of the epoch leading to the highest classification accuracy averaged over all sub-models (\emph{i.e.,} exits or switches) on validation set. 

\subsection{Results}

Results of our experiments with MSDNet are reported in Figure~\ref{fig:MSDNet_results} and Table~\ref{tab:MSDNet_results}. We may see that, as compared to the baseline supervised training without distillation, the investigated IPKD training improves the results for \mbox{CIFAR-10} dataset. The proposed IPKD-TA-1 offers the best performance for some early exits in CIFAR-100 dataset. Overall, the proposed IPKD-TA-M training approach offers better performance for both CIFAR-10 and CIFAR-100 datasets than the IPKD and the baseline. It is also interesting to see that the improvement is consistent at most intermediate classifiers.

Figure~\ref{fig:Slimmable_results} and Table~\ref{tab:Slimmable_results} summarize our experiment results with Slimmable MobileNet-V1.
The improvements of the IPKD-TA-M, as compared to the IPKD training baseline, are also consistent and convincing for the CIFAR-100 dataset.
We can see that the proposed IPKD-TA-M offers the best performance for all sub-models for CIFAR-100 dataset while resulting in similar performance with the IPKD and IPKD-TA-1 on CIFAR-10 dataset. This confirms our hypothesis that reducing big gaps between the teacher model and small student sub-models by exploiting teacher assistants is helpful.

\section{Conclusion}
\label{sec:conclusion}

In this work we have considered a family of flexible DNNs that are able instantly adapting to the available (\emph{e.g.,} computational and memory) resources for an efficient deployment. We have mainly focused on improving training of those models. Starting from recently proposed IPKD training, we have noted that this approach might be less efficient when the gap (in size) between the largest teacher model and a student sub-model is big. As such, to overcome this drawback and inspired by recently proposed knowledge distillation with teacher assistant, we have introduced new so-called {\it IPKD with teacher assistant (IPKD-TA)} flexible model training strategies.

Our proposed training strategies are general and applicable to many existing flexible DNN approaches. We have investigated them and compared to the state-of-the-art for two different flexible architectures (MSDNet and Slimmable \mbox{MobileNet-V1}) on two popular image classification benchmarks (CIFAR-10 and CIFAR-100). We have observed that in most cases one of the proposed IPKD-TA approach outperforms the state-of-the-art training methods.

\bibliographystyle{IEEEbib}
\balance
\bibliography{refs}

\end{document}